\documentclass[letterpaper, 10pt,onecolumn]{IEEEtran}

\usepackage{algorithm}
\usepackage{algpseudocode}
\usepackage{pifont}
\usepackage{multirow}
\usepackage[T1]{fontenc}
\usepackage{amsthm,xpatch}
\usepackage{amsmath,amsfonts}
\usepackage{cite}
\usepackage{amssymb}
\usepackage{dsfont}
\usepackage{color}
\usepackage{breqn}
\usepackage{mathtools}
\usepackage{bbm}
\usepackage{latexsym}
\usepackage{accents}
\usepackage{tikz}
\usepackage[mathscr]{eucal}
\usepackage{caption}
\usepackage{subcaption}
\usepackage{booktabs}
\usetikzlibrary{calc}
\usetikzlibrary{math}
\usetikzlibrary{arrows.meta}
\usepackage{csquotes}
\usepackage{varioref}

\newtheorem{example}{Example}
\DeclareMathOperator*{\argmax}{\arg\!\max}

\newcommand{\mbf}[1]{\mathbf{#1}}
\newcommand*{\transpose}{%
  {\mathpalette\@transpose{}}%
}

\IEEEoverridecommandlockouts

\begin{document}

\newcommand{\SB}[3]{
\sum_{#2 \in #1}\biggl|\overline{X}_{#2}\biggr| #3
\biggl|\bigcap_{#2 \notin #1}\overline{X}_{#2}\biggr|
}

\newcommand{\Mod}[1]{\ (\textup{mod}\ #1)}

\newcommand{\overbar}[1]{\mkern 0mu\overline{\mkern-0mu#1\mkern-8.5mu}\mkern 6mu}

\makeatletter
\newcommand*\nss[3]{%
  \begingroup
  \setbox0\hbox{$\m@th\scriptstyle\cramped{#2}$}%
  \setbox2\hbox{$\m@th\scriptstyle#3$}%
  \dimen@=\fontdimen8\textfont3
  \multiply\dimen@ by 4             
  \advance \dimen@ by \ht0
  \advance \dimen@ by -\fontdimen17\textfont2
  \@tempdima=\fontdimen5\textfont2  
  \multiply\@tempdima by 4
  \divide  \@tempdima by 5          
  \ifdim\dimen@<\@tempdima
    \ht0=0pt                        
    \@tempdima=\fontdimen5\textfont2
    \divide\@tempdima by 4          
    \advance \dimen@ by -\@tempdima 
    \ifdim\dimen@>0pt
      \@tempdima=\dp2
      \advance\@tempdima by \dimen@
      \dp2=\@tempdima
    \fi
  \fi
  #1_{\box0}^{\box2}%
  \endgroup
  }
\makeatother

\makeatletter
\renewenvironment{proof}[1][\proofname]{\par
  \pushQED{\qed}%
  \normalfont \topsep6\p@\@plus6\p@\relax
  \trivlist
  \item[\hskip\labelsep
        \itshape
    #1\@addpunct{:}]\ignorespaces
}{%
  \popQED\endtrivlist\@endpefalse
}
\makeatother

\makeatletter
\newsavebox\myboxA
\newsavebox\myboxB
\newlength\mylenA

\newcommand*\xoverline[2][0.75]{%
    \sbox{\myboxA}{$\m@th#2$}%
    \setbox\myboxB\null
    \ht\myboxB=\ht\myboxA%
    \dp\myboxB=\dp\myboxA%
    \wd\myboxB=#1\wd\myboxA
    \sbox\myboxB{$\m@th\overline{\copy\myboxB}$}
    \setlength\mylenA{\the\wd\myboxA}
    \addtolength\mylenA{-\the\wd\myboxB}%
    \ifdim\wd\myboxB<\wd\myboxA%
       \rlap{\hskip 0.5\mylenA\usebox\myboxB}{\usebox\myboxA}%
    \else
        \hskip -0.5\mylenA\rlap{\usebox\myboxA}{\hskip 0.5\mylenA\usebox\myboxB}%
    \fi}
\makeatother

\xpatchcmd{\proof}{\hskip\labelsep}{\hskip3.75\labelsep}{}{}

\pagestyle{plain}

\title{\fontsize{22.59}{28}\selectfont 
Noisy Group Testing with Side Information
}

\author{Esmaeil Karimi, Anoosheh Heidarzadeh, Krishna R. Narayanan, and Alex Sprintson\thanks{The authors are with the Department of Electrical and Computer Engineering, Texas A\&M University, College Station, TX 77843 USA (E-mail: \{esmaeil.karimi, anoosheh, krn, spalex\}@tamu.edu).}}


\maketitle 

\thispagestyle{plain}

\begin{abstract}
Group testing has recently attracted significant attention from the research community due to its applications in diagnostic virology. An instance of the group testing problem includes a ground set of individuals which includes a small subset of infected individuals. The group testing procedure consists of a number of tests, such that each test indicates whether or not a given subset of individuals includes one or more infected individuals. The goal of the group testing procedure is to identify the subset of infected individuals with the minimum number of tests.
Motivated by practical scenarios, such as testing for viral diseases, this paper focuses on the following group testing settings: (i) the group testing procedure is noisy, i.e., the outcome of the group testing procedure can be flipped with a certain probability; (ii) there is a certain amount of side information on the distribution of the infected individuals available to the group testing algorithm. The paper makes the following contributions. First, we propose a probabilistic model, referred to as an \emph{interaction model}, that captures the side information about the probability distribution of the infected individuals. Next, we present a decoding scheme, based on the belief propagation, that leverages the interaction model to improve the decoding accuracy. Our results indicate that the proposed algorithm achieves higher success probability and lower false-negative and false-positive rates when compared to the traditional belief propagation especially in the high noise regime.
\end{abstract}

\section{introduction}
Identifying infected people is a critical step in dealing with pandemics caused by viral diseases. However, testing a large number of people individually might be prohibitively expensive for practical reasons. For this reason, we need to deploy strategies that allow efficient testing.
 \emph{Group Testing} (GT) has been shown as an efficient strategy in reducing the number of tests required to test for pandemics.
An instance of the GT problem includes a  set $\mbf{S}$ of $N$ individuals which includes a small subset of infected individuals. The GT procedure consists of a sequence of tests, such that each test indicates whether there are one or more infected individuals in a given subset of $\mbf{S}$. The goal of the GT procedure is to identify the subset of infected individuals through the minimum number of tests. 

The GT problem has been the subject of many studies. Most studies have focused on the following two models \cite{8989403,8635972,DBLP:journals/corr/abs-2110-10110,MCDERMOTT2021532}: (i) a  \emph{combinatorial} model which assumes that the number of infected individuals is fixed and known; (ii) a \emph{probabilistic} model which assumes that each individual is infected with a certain probability. There are also two types of GT algorithms:  \emph{non-adaptive}, and \emph{adaptive}. 
In this paper, we are interested in {non-adaptive} GT strategies, where all tests are designed in advance. This is in contrast to {adaptive} strategies, in which the design of each test depends on the results of the previous tests~\cite{9646274,8437774,8919896,CIT-099}. 

Motivated by practical scenarios where the outcome of the tests can be affected by noise, we focus on the \emph{noisy GT} setting, in which the outcome of a test can be flipped with some probability. In the noisy GT setting, the goal is to identify the set of infected individuals with high probability $(1-\varepsilon)$, for small values of $\varepsilon$. We also focus on a variation of a probabilistic GT model in which the prior infection probability is not uniform and in which there is a certain amount of side information on the distribution of the infected individuals available to the GT algorithm. 

A GT algorithm consists of two parts: \emph{encoding} and \emph{decoding}. The encoding part is concerned with the test design, i.e., the decision on which individuals to include in each test. The decoding part is concerned with identifying the infected individuals given the test design and outcomes of the tests. Different decoding algorithms such as linear programming, combinatorial orthogonal matching pursuit, definite defectives, belief propagation (BP), and separate decoding of items have been proposed for noisy non-adaptive GT. A thorough review and comparison of these algorithms is provided in~\cite{CIT-099}.
In the context of GT, it is extremely difficult to analyze the performance of BP algorithms even for the asymptotic regime. To the best of our knowledge, no theoretical analysis has been provided for the BP-based GT algorithms so far.
However, empirical evidence suggests that the BP algorithm results in lower error probabilities compared to other algorithms for the probabilistic model.
BP is a message passing algorithm that passes messages over the edges in the underlying factor graph representation of the GT problem. For a cycle-free factor graph, BP decoding is equivalent to Maximum a Posteriori (MAP) decoding. However, in the presence of loops in the factor graph, BP becomes suboptimal.

This paper focuses on leveraging the side information for improving the performance of BP-based decoding algorithms for noisy GT. In the context of testing for viral infections, different forms of side information can be exploited including the prevalence rate, individuals' symptoms, family structure, community structure, and contact tracing information. It has been shown that side information can be used to reduce the required tests~\cite{zhu2020noisy,DBLP:journals/corr/abs-2007-08111,DBLP:journals/corr/abs-2012-02804,DBLP:journals/corr/abs-2106-02699,DBLP:journals/corr/abs-2101-02405}. 
   For example, Zhu \emph{et al.}~\cite{zhu2020noisy} show that the number of tests can be reduced if the prior information about the prevalence rate is takend into account. Nikolopoulos \emph{et al.}~\cite{DBLP:journals/corr/abs-2007-08111,DBLP:journals/corr/abs-2012-02804} and Ahn \emph{et al.}~\cite{DBLP:journals/corr/abs-2101-02405} show that utilizing community structure also leads to a lower number of tests. 
While the focus of these works is on the encoder design, in our work we focus on leveraging the side information for the efficient decoder design.

\subsection{Contribution.}

In this work, first, we propose a probabilistic model, referred to as an \emph{interaction model}, that captures the side information about the probability distribution of the infected individuals. Our model is motivated by the availability of contact tracing information which can be collected from surveys and mobile phone applications~\cite{KleinmanE653,doi:10.1080/19371918.2020.1806170,munzert2021tracking}. 
 Next, we present a decoding scheme, based on belief propagation, that leverages the interaction model to improve the decoding accuracy. Our results indicate that the proposed algorithm achieves higher success probability and lower false-negative and false-positive rates when compared to the traditional belief propagation especially in the high noise regime.

\section{Probabilistic Model}\label{sec:SN}
Throughout the paper, we denote vectors and matrices by bold-face small and capital letters, respectively. For an integer $i\geq 1$, we denote $\{1,\dots,i\}$ by $[i]$. Let $\mbf{S}$ be a set. The set of all subsets of size $\ell$ for set $\mbf{S}$ is denoted by $\{\mbf{S}\}_\ell$.

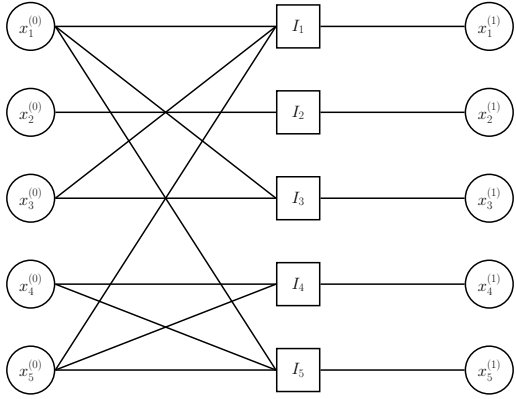
\begin{figure}
\centering
\begin{tikzpicture}
\def\horzgap{2in}; 
\def \gapVN{0.45in}; 
\def \gapCN{0.45in}; 

\def\nodewidth{0.25in};
\def\nodewidthA{0.3in};
\def \edgewidth{0.01in};
\def\ext{0.1in};
\def \dotwidth{0.3mm};

\tikzstyle{dots} = [rectangle, draw,line width=0.2mm,  inner sep=0mm, fill=black, minimum height=\dotwidth, minimum width=\dotwidth]
\tikzstyle{check} = [rectangle, draw,line width=0.2mm,  inner sep=0mm, fill=white, minimum height=0.75*\nodewidthA, minimum width=0.75*\nodewidthA]
\tikzstyle{check1} = [rectangle, draw, dotted,line width=0.4mm,  inner sep=0mm, fill=white, minimum height=0.6*\nodewidthA, minimum width=0.6*\nodewidthA]
\tikzstyle{bit0} = [circle, draw, line width=0.2mm, inner sep=0mm, fill=white, minimum size=\nodewidth]
\tikzstyle{bit1} = [circle, draw, line width=0.2mm, inner sep=0mm, fill=blue, minimum size=\nodewidth]
\tikzstyle{bituncover} = [circle, draw=none, line width=0.2mm, inner sep=0mm, fill=gray, minimum size=\nodewidthA]
\tikzstyle{edgesock} = [circle, inner sep=0mm, minimum size=\edgewidth,draw, fill=white]

\foreach \vn in {1,2,3,4,5}{
 \node[bit0] (vn0\vn) at (-0.7*\horzgap,-\vn*\gapVN) {};
}                    
\path (vn01) ++(0,0) node()[scale=0.25, inner sep=0mm] {\Huge{$x_{1}^{(0)}$}};
\path (vn02) ++(0,0) node()[scale=0.25, inner sep=0mm] {\Huge{$x_{2}^{(0)}$}};
\path (vn03) ++(0,0) node()[scale=0.25, inner sep=0mm] {\Huge{$x_{3}^{(0)}$}};
\path (vn04) ++(0,0) node()[scale=0.25, inner sep=0mm] {\Huge{$x_{4}^{(0)}$}};
\path (vn05) ++(0,0) node()[scale=0.25, inner sep=0mm] {\Huge{$x_{5}^{(0)}$}};

\foreach \vn in {1,2,3,4,5}{
 \node[bit0] (vn1\vn) at (0.5*\horzgap,-\vn*\gapVN) {};
}                    
\path (vn11) ++(0,0) node()[scale=0.25, inner sep=0mm] {\Huge{$x_{1}^{(1)}$}};
\path (vn12) ++(0,0) node()[scale=0.25, inner sep=0mm] {\Huge{$x_{2}^{(1)}$}};
\path (vn13) ++(0,0) node()[scale=0.25, inner sep=0mm] {\Huge{$x_{3}^{(1)}$}};
\path (vn14) ++(0,0) node()[scale=0.25, inner sep=0mm] {\Huge{$x_{4}^{(1)}$}};
\path (vn15) ++(0,0) node()[scale=0.25, inner sep=0mm] {\Huge{$x_{5}^{(1)}$}};

\foreach \cn in {1,...,5}{
\node[check] (cn\cn) at (0,-\cn*\gapVN) {};
}
\draw[line width=0.2mm] (vn01.east)--(cn1.west);
\draw[line width=0.2mm] (vn03.east)--(cn1.west);
\draw[line width=0.2mm] (vn05.east)--(cn1.west);

\draw[line width=0.2mm] (vn02.east)--(cn2.west);

\draw[line width=0.2mm] (vn01.east)--(cn3.west);
\draw[line width=0.2mm] (vn03.east)--(cn3.west);

\draw[line width=0.2mm] (vn04.east)--(cn4.west);
\draw[line width=0.2mm] (vn05.east)--(cn4.west);

\draw[line width=0.2mm] (vn01.east)--(cn5.west);
\draw[line width=0.2mm] (vn04.east)--(cn5.west);
\draw[line width=0.2mm] (vn05.east)--(cn5.west);

\draw[line width=0.2mm] (cn1.east)--(vn11.west);
\draw[line width=0.2mm] (cn2.east)--(vn12.west);
\draw[line width=0.2mm] (cn3.east)--(vn13.west);
\draw[line width=0.2mm] (cn4.east)--(vn14.west);
\draw[line width=0.2mm] (cn5.east)--(vn15.west);

\def\moveX {3.2*\nodewidth};
\def\moveXA {2*\nodewidth};

\path (cn1)++(0,0) node ()[scale=0.25] {\Huge{$I_1$}};
    \path (cn2)++(0,0) node ()[scale=0.25] {\Huge{$I_2$}};
    \path (cn3)++(0,0) node ()[scale=0.25] {\Huge{$I_3$}};
    \path (cn4)++(0,0) node ()[scale=0.25] {\Huge{$I_4$}};
     \path (cn5)++(0,0) node ()[scale=0.25] {\Huge{$I_5$}};

\end{tikzpicture}
\caption{An example of an interaction graph. Nodes $x_i^{(0)}$ and  $x_i^{(1)}$ represent individual $i$ at times 0 and 1, respectively. An interaction node $I_i$ captures interactions between individual $i$ and other individuals from time $0$ to time $1$.} \label{fig:Interaction}
\end{figure}

Our model assumes that there are two points in time, namely time $0$ and time $1$ such that time $0$ occurs prior to time $1$. Let $N$ be the total number of individuals. 
We define the vector $\mathbf{x}^{(0)}\in \{0,1\}^N$ to represent the status of $N$ individuals at time $0$, such that $x_i^{(0)}$ is $1$ if the $i$-th individual is infected at time $0$, and is $0$ otherwise. 
Similarly, we define the vector $\mathbf{x}^{(1)}\in \{0,1\}^N$ to represent the status of $N$ individuals at time $1$. We assume that at time 0, the probability of an individual being infected is equal to the prevalence rate $p$, and that the infection of each individual at time $0$ occurs independently of other individuals. The probability of the individual to be infected at time 1 depends on their probability to be infected at time 0 as well as their interaction with other individuals. The interactions of individuals between time 0 and time 1 is captured by the \emph{interaction graph}. For each individual $i$, the graph includes nodes $x_i^{(0)}$ and  $x_i^{(1)}$ that represent that individual at times 0 and 1, respectively. For each individual $i$, the graph has an \emph{interaction node} $I_i$  that captures interactions between individual $i$ and other individuals from time $0$ to time $1$.  In particular, the graph contains an edge $(x_j^{(0)},I_i)$  for each individual $j$ who have been in contact with individual $i$ from time $0$ to time $1$. An example of an interaction graph is shown in Fig.~\ref{fig:Interaction}.

We assume that an infected individual infects a healthy individual with probability $q$, referred to as \emph{contagion probability.} It is also assumed that if an individual is infected at time $0$, they remain infected by time $1$. The interaction model can be used to find the probability that an individual at time $1$ is infected. The $i$th individual is not infected at time $1$ if the following holds: $1)$ the $i$th individual is not infected at time $0$, and $2)$ other individuals in contact with the $i$th individual either are not infected at time $0$ or, if infected, they do not infect the $i$th individual. Thus, the probability of individual $i$ to be infected at time 1 can be calculated as follows:
\[ 
P\Big(x_i^{(1)}=0\Big) =(1-p)\Big(1-p+p(1-q)\Big)^{d_i}
=(1-p)(1-pq)^{d_i}, 
 \]
where $d_i$ is the number of individuals which interact with individual $i$ from time $0$ to time $1$. The probability that individual $i$ is infected at time $1$ is given by
\begin{equation}\label{eq:prior}
\pi_i=P\Big(x_i^{(1)}=1\Big)=1-P\Big(x_i^{(1)}=0\Big) =1-(1-p)(1-pq)^{d_i}. 
\end{equation}
Our ultimate goal is to test and identify infected individuals at time $1$, assuming the knowledge of the probabilistic model described above. This model can be extended to capture interactions between individuals in more than one round. For ease of exposition, we consider only one round of interactions in this study.

In non-adaptive group testing, designing a testing scheme consisting of $M$ tests is equivalent to the construction of a binary matrix with $M$ rows which is referred to as a testing matrix.
We let matrix ${\textbf{A}\in \{0,1\}^{M\times N}}$ denote the testing matrix. The entry $(t,i)$ of matrix ${\textbf{A}}$ is denoted by $a_{t,i}$.
If $a_{t,i}=1$, it means that the $i$-th item is present in the $t$-th test. 
The design of testing matrices for group testing has been studied extensively (see e.g., \cite{CIT-099}).
Our proposed algorithms are applicable for any testing matrix; however, simulation results are presented for Bernoulli designs. In a Bernoulli design, each individual is included in each test independently at random with some fixed probability $\nu/K$ where $K$ is the average number of defective items and $\nu$ is a constant.

 The standard noiseless group testing is formulated component-wise using the Boolean OR operation as ${y_t=\bigvee_{i=1}^{N} a_{t,i}x_i^{(1)}}$ where $y_t$ and $\bigvee$ are the $t$th test result and a Boolean OR operation, respectively. In this paper, we consider the widely-adopted binary symmetric noise model where the values $\bigvee_{i=1}^{N} a_{t,i}x_i^{(1)}$ are flipped independently at random with a given probability. The $t$th test result in a binary symmetric noise model is given by
\[   
     y_t=\begin{cases}
       \bigvee_{i=1}^{N} a_{t,i}x_i^{(1)} &\quad\text{with probability}~ 1-\rho,\\
        1\oplus \bigvee_{i=1}^{N} a_{t,i}x_i^{(1)} &\quad \text{with probability} ~\rho,\\
       
     \end{cases}
\] 
where $\oplus$ is the XOR operation and $\rho$ is the probability that the values $\bigvee_{i=1}^{N} a_{t,i}x_i^{(1)}$ are flipped.
Note that this model and the proposed algorithms can be easily extended to include the general binary noise model where the values $\bigvee_{i=1}^{N} a_{t,i}x_i^{(1)}$ are flipped from $0$ to $1$ and from $1$ to $0$ with different probabilities independently.
However, for ease of exposition, we focus only on the binary symmetric noise model. 
 We let vector $\mathbf{y}\in\{0,1\}^M$ denote the outcomes of the $M$ tests. 
 
 Our objective is to design a decoding algorithm that performs well under the following three metrics: (i) success probability which captures the probability that all infected individuals are identified correctly; (ii) False-Negative Rate (FNR), which is the number of infected individuals falsely classified as healthy over the total number of infected individuals, (iii) False-Positive Rate (FPR), defined as the ratio of the number of healthy individuals falsely classified as infected and the total number of healthy individuals.

\section{Proposed Decoding Algorithms \label{sec:main results}}
In this section, we describe the proposed decoding algorithms for retrieving the status vector $\mathbf{x}^{(1)}$ from the test results vector $\mathbf{y}$ and the testing matrix ${\textbf{A}}$. 
\subsection{Belief Propagation Using Initial Prior Probabilities}\label{subsec:bp}
Message passing algorithms are utilized to solve inference problems, optimization problems, and constraint satisfaction problems. In an inference problem, there are some noisy measurements as input, and the goal is to infer the value of some unobserved variables from those measurements. It is impossible, in general, to make those inferences with complete certainty, but one can try to obtain the most probable value of the unobserved variables\cite{mackay2003information,russell2003artificial}. In a probabilistic noisy group testing, we intend to perform a Maximum a Posteriori (MAP) estimation to find the status vector $\hat{\mbf{x}}^{(1)}$ given the test results vector $\mbf{u}$.
\begin{equation}\label{eq:cost}
\argmax_{\hat{\mbf{x}}^{(1)}} P\Big(\mbf{x}^{(1)}=\hat{\mbf{x}}^{(1)}\Big)P\Big(\mbf{y}=\mbf{u}\vert \mbf{x}^{(1)}=\hat{\mbf{x}}^{(1)}\Big)
\end{equation}
This problem can be solved using exhaustive search when the vector $\mbf{x}^{(1)}$ is small, e.g., $N\approx 10-20$. 
However, the exhaustive search approach rapidly becomes intractable when $N$ increases. For instance, when $N=100$ (a relatively small problem), the number of different configurations for the vector $\mbf{x}^{(1)}$ is $2^{100}$. 

\begin{figure}
\centering
\begin{tikzpicture}
\def\horzgap{2in}; 
\def \gapVN{0.45in}; 
\def \gapCN{0.45in}; 

\def\nodewidth{0.25in};
\def\nodewidthA{0.3in};
\def \edgewidth{0.01in};
\def\ext{0.1in};
\def \dotwidth{0.3mm};

\tikzstyle{dots} = [rectangle, draw,line width=0.2mm,  inner sep=0mm, fill=black, minimum height=\dotwidth, minimum width=\dotwidth]
\tikzstyle{check} = [rectangle, draw,line width=0.2mm,  inner sep=0mm, fill=white, minimum height=0.75*\nodewidthA, minimum width=0.75*\nodewidthA]
\tikzstyle{check1} = [rectangle, draw, dotted,line width=0.4mm,  inner sep=0mm, fill=white, minimum height=0.6*\nodewidthA, minimum width=0.6*\nodewidthA]
\tikzstyle{bit0} = [circle, draw, line width=0.2mm, inner sep=0mm, fill=white, minimum size=\nodewidth]
\tikzstyle{bit1} = [circle, draw, line width=0.2mm, inner sep=0mm, fill=blue, minimum size=\nodewidth]
\tikzstyle{bituncover} = [circle, draw=none, line width=0.2mm, inner sep=0mm, fill=gray, minimum size=\nodewidthA]
\tikzstyle{edgesock} = [circle, inner sep=0mm, minimum size=\edgewidth,draw, fill=white]

\foreach \vn in {1,2,3,4,5}{
 \node[bit0] (vn\vn) at (0,-\vn*\gapVN) {};
}                    
\path (vn1) ++(0,0) node()[scale=0.25, inner sep=0mm] {\Huge{$x_{1}^{(1)}$}};
\path (vn2) ++(0,0) node()[scale=0.25, inner sep=0mm] {\Huge{$x_{2}^{(1)}$}};
\path (vn3) ++(0,0) node()[scale=0.25, inner sep=0mm] {\Huge{$x_{3}^{(1)}$}};
\path (vn4) ++(0,0) node()[scale=0.25, inner sep=0mm] {\Huge{$x_{4}^{(1)}$}};
\path (vn5) ++(0,0) node()[scale=0.25, inner sep=0mm] {\Huge{$x_{5}^{(1)}$}};
\foreach \cn in {1,...,3}{
\node[check] (cn\cn) at (0.7*\horzgap,-\cn*\gapCN-0.4in) {};
}
\draw[line width=0.2mm] (vn1.east)--(cn1.west);
\draw[line width=0.2mm] (vn3.east)--(cn1.west);
\draw[line width=0.2mm] (vn5.east)--(cn1.west);
\draw[line width=0.2mm] (vn2.east)--(cn2.west);
\draw[line width=0.2mm] (vn3.east)--(cn2.west);
\draw[line width=0.2mm] (vn4.east)--(cn2.west);
\draw[line width=0.2mm] (vn2.east)--(cn3.west);
\draw[line width=0.2mm] (vn4.east)--(cn3.west);
\draw[line width=0.2mm] (vn5.east)--(cn3.west);
\foreach \pn in {1,...,5}{\node[check1] (pn\pn) at (-0.7in,-\pn*\gapVN) {};}
\draw[dotted, line width=0.4mm] (pn1.east)--(vn1.west);
\draw[dotted, line width=0.4mm] (pn2.east)--(vn2.west);
\draw[dotted, line width=0.4mm] (pn3.east)--(vn3.west);
\draw[dotted, line width=0.4mm] (pn4.east)--(vn4.west);
\draw[dotted, line width=0.4mm] (pn5.east)--(vn5.west);
\def\moveX {3.2*\nodewidth};
\def\moveXA {2*\nodewidth};
\path (cn1)++(0,0) node ()[scale=0.25] {\Huge{$y_1$}};
    \path (cn2)++(0,0) node ()[scale=0.25] {\Huge{$y_2$}};
    \path (cn3)++(0,0) node ()[scale=0.25] {\Huge{$y_3$}};
\path (pn1.north)++(0,0.5) node ()[scale=0.3] {\Huge{Priors}};       
\path (vn1.north)++(0,0.5) node ()[scale=0.3] {\Huge{Individuals}};   
\path (cn1.north)++(0,0.5) node ()[scale=0.3] {\Huge{Tests}};
\end{tikzpicture}
\caption{An example of a factor graph representing a group testing scheme. A variable node $x_{i}^{(1)}$ represents individual $i$ at time $1$. A factor node $y_t$ represents test $t$. Factor nodes represented by dotted squares correspond to the \textit{a priori} probability distribution of the variable nodes. } \label{fig:Bipartite}
\end{figure}
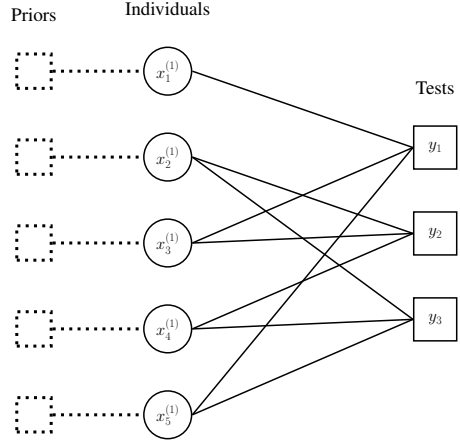

An alternative solution is to find the marginals of the posterior distribution for each item using Belief Propagation (BP). BP is a message passing algorithm for performing inference on factor graphs with the purpose of calculating the marginal distribution for each unobserved variable, conditional on observed variables. A factor graph is a type of probabilistic graphical model which is used to visualize and precisely define the underlying optimization problem. Factor graphs are bipartite graphs with two types of nodes referred to as variable nodes and factor nodes. See Fig.~\ref{fig:Bipartite} for an example of a factor graph.  The variable nodes which represent the variables in the optimization problem are represented by circles. The factor nodes show how the overall cost function can be factorized into local cost functions and are represented by squares. There is an edge between the variables that are involved in a local cost function and the factor node representing that local cost function.

We assume that the only side information we have is the prevalence rate at time $0$. Since we have no information about the status of individuals at time $1$, we consider the prior probability of each individual being infected at time $1$ to be equal to the prevalence rate $p$, independent of other individuals. The overall cost function in~\eqref{eq:cost} can be factorized as follows.
\begin{equation}
 P\Big(\mbf{x}^{(1)}=\hat{\mbf{x}}^{(1)}\Big)P\Big(\mbf{y}=\mbf{u}\vert \mbf{x}^{(1)}=\hat{\mbf{x}}^{(1)}\Big) =\Bigg[\prod_{i=1}^N P\Big({x}_i^{(1)}=\hat{{x}}_i^{(1)}\Big)\Bigg] \times \Bigg[\prod_{t=1}^M P\Big(y_t=u_t \Big \vert \Big\{{{x}}_{i}^{(1)}=\hat{{x}}_{i}^{(1)}\Big\}_{i\in \mathcal{N}(t)}\Big)\Bigg], 
\end{equation}
where $\mathcal{N}(t)$ denotes the indices of individuals involved in the $t$th test. 

In order to apply BP, we consider the factor graph (Tanner graph) representation of the group testing scheme. In the Tanner graph, there are $N$ variable nodes that represent individuals at time $1$. There are also $M$ factor nodes that represent the tests. Each test factor node corresponds to the conditional probability distribution of a test result, given the observed variable nodes. 
Each individual in the Tanner graph is connected to the test in which the individual participates, according to the testing matrix. There are also $N$ factor nodes that correspond to the \textit{a priori} probability distribution of the variable nodes. Since these factor nodes are usually not exhibited in a Tanner graph, we show them using dotted squares in Fig.~\ref{fig:Bipartite}.

For a cycle-free factor graph, BP decoding is equivalent to Maximum a Posteriori (MAP) decoding. However, in the presence of loops in the factor graph, BP becomes suboptimal. In other words, loopy BP computes an approximation of the marginals of the posterior distribution for each variable node. For a loopy BP algorithm, the messages are passed iteratively from variable nodes to factor nodes and vice versa. We let ${\mu}_{i \rightarrow t} = [\mu_{i\rightarrow t}(0) \ \mu_{i\rightarrow t}(1)]$ and ${\mu}_{t\rightarrow i} = [\mu_{t \rightarrow i}(0) \ \mu_{t\rightarrow i}(1)]$ denote the message from individual $i$ to test $t$ and the message from test $t$ to individual $i$, respectively. In general, the message from variable node $i$ to factor node $t$ is given by computing the product of all incoming messages from the neighboring factor nodes of variable node $i$ excluding the message from factor node $t$.
\begin{equation}\label{eq:itm-to-tst}
\begin{cases}
 \mu_{i\rightarrow t}(0) \propto (1-p) \displaystyle \prod_{t^{\prime}\in \mathcal{N}(i)\setminus \{t\}}\mu_{t^{\prime}\rightarrow i}(0), \\
 \mu_{i\rightarrow t}(1) \propto p \displaystyle \prod_{t^{\prime}\in \mathcal{N}(i)\setminus \{t\}}\mu_{t^{\prime}\rightarrow i}(1),\\
  \end{cases}
\end{equation}
where $\propto$ indicates equality up to a
normalizing constant, and $\mathcal{N}(i)$ denotes the indices of tests in which item $i$ participates. Note that these messages are probability distributions on \{0,1\}, i.e., $\mu_{i\rightarrow t}(0) + \mu_{i\rightarrow t}(1)=1 $. 
Since we assume that the prior probability of each individual being infected at time $1$ is equal to $p$, the messages are initialized by
\begin{equation}\label{eq:intl}
 \mu_{i\rightarrow t}(1)=1-\mu_{i\rightarrow t}(0)=p.   
\end{equation}
The messages from factor nodes to variable nodes are computed as follows. The message from test $t$ to individual $i$ is given by
\begin{equation}\label{eq:ctov}
 \mu_{t\rightarrow i}(\hat{{x}}_{i}^{(1)})=  \sum_{\substack{\hat{{x}}_{i^\prime}^{(1)}\in\{0,1\},~{i^{\prime}\in \mathcal{N}(t)\setminus\{i\}}\\u_t\in \{0,1\}}} \Bigg[P\Big(y_t=u_t\Big\vert {x}_{i}^{(1)}=\hat{{x}}_{i}^{(1)},\Big\{{x}_{i^\prime}^{(1)}=\hat{{x}}_{i^\prime}^{(1)}\Big\}_{i^{\prime}\in \mathcal{N}(t)\setminus\{i\}}\Big) \prod_{i^{\prime}\in \mathcal{N}(t)\setminus\{i\}}\mu_{i^{\prime}\rightarrow t}\Big(\hat{{x}}_{i^\prime}^{(1)}\Big)\Bigg].
\end{equation}
As was shown in~\cite{barbier2020strong,2018,coja2018belief,doi:10.1080/00018732.2016.1211393}, the equation~\eqref{eq:ctov} can be simplified as follows. If $y_t=0$, we have 
\begin{equation}\label{eq:tst-to-itm0}
\begin{cases}
 \mu_{t\rightarrow i}(0) \propto  \rho + (1-2\rho) \displaystyle \prod_{i^{\prime}\in \mathcal{N}(t)\setminus \{i\}}\mu_{i^{\prime}\rightarrow t}(0), \\
 \mu_{t\rightarrow i}(1) \propto  \rho, \\
  \end{cases}
\end{equation} and if $y_t=1$, we have
\begin{equation}\label{eq:tst-to-itm1}
\begin{cases}
 \mu_{t\rightarrow i}(0) \propto  1-\rho - (1-2\rho) \displaystyle \prod_{i^{\prime}\in \mathcal{N}(t)\setminus \{i\}}\mu_{i^{\prime}\rightarrow t}(0), \\
 \mu_{t\rightarrow i}(1) \propto  1-\rho. \\
  \end{cases}
\end{equation}

We perform a fixed point iteration using the BP equations~\eqref{eq:itm-to-tst},~\eqref{eq:tst-to-itm0}, and~\eqref{eq:tst-to-itm1}. We stop the algorithm after a fixed number $T$ of iterations. The parameter $T$ is chosen experimentally. In the end, we compute the marginals of the posterior distribution for each variable node by computing the product of all incoming messages from the neighboring factor nodes of that variable node.

\begin{equation}
\begin{cases}
{{\phi}}\Big(x_{i}^{(1)}=0\Big) \propto  (1-p) \displaystyle \prod_{t\in \mathcal{N}(i)}\mu_{t\rightarrow i}(0), \\
  {{\phi}}\Big(x_{i}^{(1)}=1\Big) \propto  p \displaystyle \prod_{t\in \mathcal{N}(i)}\mu_{t\rightarrow i}(1). \\
  \end{cases} \vspace{0.1cm}
\end{equation}
For convenience, we work with the Log-Likelihood Ratio (LLR) of a marginal defined as
\begin{equation}\label{eq:llr}
\lambda_i=\ln \frac{ {{\phi}}\Big(x_{i}^{(1)}=1\Big)}{{{\phi}}\Big(x_{i}^{(1)}=0\Big)}=\ln \frac{p}{1-p} + \sum_{t\in \mathcal{N}(i)}\ln \frac{\mu_{t\rightarrow i}(1)}{\mu_{t\rightarrow i}(0)}.
\end{equation}

We consider a threshold, $\tau$, and announce the $i$th individual infected if $ \lambda_i \geq \tau$. A natural threshold one can choose is $\tau=0$. Note that values other than $0$ are also permissible. 
Algorithm~\ref{alg:BP} defines the belief propagation using initial prior probabilities algorithm.

\begin{algorithm}
\caption{Belief Propagation Using Initial Prior Probabilities}
\label{alg:BP}
\label{CHalgorithm}
\begin{algorithmic}[1]
\State Initialize $\mu_{i\rightarrow t}(1)=1-\mu_{i\rightarrow t}(0)=p$ $\forall i\in [N], \forall t\in \mathcal{N}(i)$
\For{$\ell=1,2,\cdots,T$ }
\State Compute $\mu_{t\rightarrow i}(0)$ and $\mu_{t\rightarrow i}(1)$ $\forall t\in [M], \forall i\in \mathcal{N}(t)$ using~\eqref{eq:tst-to-itm0} and~\eqref{eq:tst-to-itm1}
\State Compute $\mu_{i\rightarrow t}(0)$ and $\mu_{i\rightarrow t}(1)$ $\forall i\in [N], \forall t\in \mathcal{N}(i)$ using~\eqref{eq:itm-to-tst}
\EndFor
\State Compute $\lambda_i$ $\forall  i\in [N]$ using~\eqref{eq:llr}
\end{algorithmic}
\end{algorithm}

\subsection{Belief Propagation Using Updated Prior Probabilities.}

In this scheme, instead of using the prevalence rate at time $0$ for the probability that an individual is infected at time $1$, we use the updated prior probability $\pi_i$, $i\in [N]$, given by~\eqref{eq:prior}. We perform the BP algorithm in Section~\ref{subsec:bp} where in the equations~\eqref{eq:itm-to-tst}-\eqref{eq:llr}, the initial prior probability $p$ is replaced by the updated prior probability $\pi_i$, for each $i\in [N]$.

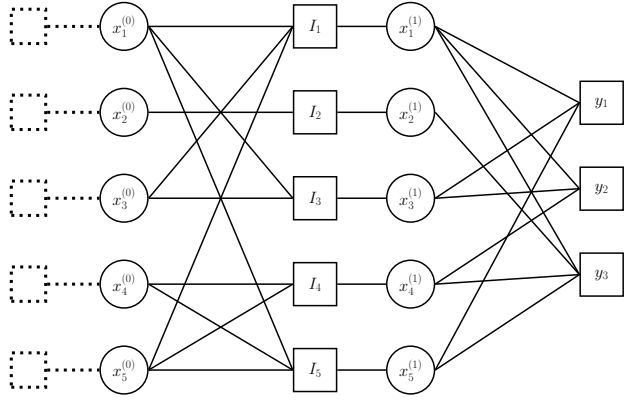
\begin{figure}
\centering
\begin{tikzpicture}
\def\horzgap{2in}; 
\def \gapVN{0.45in}; 
\def \gapCN{0.45in}; 

\def\nodewidth{0.25in};
\def\nodewidthA{0.3in};
\def \edgewidth{0.01in};
\def\ext{0.1in};
\def \dotwidth{0.3mm};

\tikzstyle{dots} = [rectangle, draw,line width=0.2mm,  inner sep=0mm, fill=black, minimum height=\dotwidth, minimum width=\dotwidth]
\tikzstyle{check} = [rectangle, draw,line width=0.2mm,  inner sep=0mm, fill=white, minimum height=0.75*\nodewidthA, minimum width=0.75*\nodewidthA]
\tikzstyle{check1} = [rectangle, draw, dotted,line width=0.4mm,  inner sep=0mm, fill=white, minimum height=0.6*\nodewidthA, minimum width=0.6*\nodewidthA]
\tikzstyle{bit0} = [circle, draw, line width=0.2mm, inner sep=0mm, fill=white, minimum size=\nodewidth]
\tikzstyle{bit1} = [circle, draw, line width=0.2mm, inner sep=0mm, fill=blue, minimum size=\nodewidth]
\tikzstyle{bituncover} = [circle, draw=none, line width=0.2mm, inner sep=0mm, fill=gray, minimum size=\nodewidthA]

\tikzstyle{edgesock} = [circle, inner sep=0mm, minimum size=\edgewidth,draw, fill=white]     

\foreach \vn in {1,2,3,4,5}{
 \node[bit0] (vn0\vn) at (-0.65*\horzgap,-\vn*\gapVN) {};
}                    
\path (vn01) ++(0,0) node()[scale=0.25, inner sep=0mm] {\Huge{$x_{1}^{(0)}$}};
\path (vn02) ++(0,0) node()[scale=0.25, inner sep=0mm] {\Huge{$x_{2}^{(0)}$}};
\path (vn03) ++(0,0) node()[scale=0.25, inner sep=0mm] {\Huge{$x_{3}^{(0)}$}};
\path (vn04) ++(0,0) node()[scale=0.25, inner sep=0mm] {\Huge{$x_{4}^{(0)}$}};
\path (vn05) ++(0,0) node()[scale=0.25, inner sep=0mm] {\Huge{$x_{5}^{(0)}$}};

\foreach \vn in {1,2,3,4,5}{
 \node[bit0] (vn1\vn) at (0.1*\horzgap,-\vn*\gapVN) {};
}                    
\path (vn11) ++(0,0) node()[scale=0.25, inner sep=0mm] {\Huge{$x_{1}^{(1)}$}};
\path (vn12) ++(0,0) node()[scale=0.25, inner sep=0mm] {\Huge{$x_{2}^{(1)}$}};
\path (vn13) ++(0,0) node()[scale=0.25, inner sep=0mm] {\Huge{$x_{3}^{(1)}$}};
\path (vn14) ++(0,0) node()[scale=0.25, inner sep=0mm] {\Huge{$x_{4}^{(1)}$}};
\path (vn15) ++(0,0) node()[scale=0.25, inner sep=0mm] {\Huge{$x_{5}^{(1)}$}};

\foreach \cn in {1,...,5}{
\node[check] (cn\cn) at (-0.15*\horzgap,-\cn*\gapVN) {};
}
\draw[line width=0.2mm] (vn01.east)--(cn1.west);
\draw[line width=0.2mm] (vn03.east)--(cn1.west);
\draw[line width=0.2mm] (vn05.east)--(cn1.west);

\draw[line width=0.2mm] (vn02.east)--(cn2.west);

\draw[line width=0.2mm] (vn01.east)--(cn3.west);
\draw[line width=0.2mm] (vn03.east)--(cn3.west);

\draw[line width=0.2mm] (vn04.east)--(cn4.west);
\draw[line width=0.2mm] (vn05.east)--(cn4.west);

\draw[line width=0.2mm] (vn01.east)--(cn5.west);
\draw[line width=0.2mm] (vn04.east)--(cn5.west);
\draw[line width=0.2mm] (vn05.east)--(cn5.west);

\draw[line width=0.2mm] (cn1.east)--(vn11.west);
\draw[line width=0.2mm] (cn2.east)--(vn12.west);
\draw[line width=0.2mm] (cn3.east)--(vn13.west);
\draw[line width=0.2mm] (cn4.east)--(vn14.west);
\draw[line width=0.2mm] (cn5.east)--(vn15.west);

\foreach \cn in {1,...,3}{
\node[check] (cn1\cn) at (0.6*\horzgap,-\cn*\gapCN-0.4in) {};
}

\draw[line width=0.2mm] (vn11.east)--(cn11.west);
\draw[line width=0.2mm] (vn13.east)--(cn11.west);
\draw[line width=0.2mm] (vn15.east)--(cn11.west);

\draw[line width=0.2mm] (vn13.east)--(cn12.west);
\draw[line width=0.2mm] (vn14.east)--(cn12.west);
\draw[line width=0.2mm] (vn11.east)--(cn12.west);

\draw[line width=0.2mm] (vn12.east)--(cn13.west);
\draw[line width=0.2mm] (vn14.east)--(cn13.west);
\draw[line width=0.2mm] (vn15.east)--(cn13.west);
\draw[line width=0.2mm] (vn11.east)--(cn13.west);

\foreach \pn in {1,...,5}{\node[check1] (pn\pn) at (-0.9*\horzgap,-\pn*\gapVN) {};}
\draw[dotted, line width=0.4mm] (pn1.east)--(vn01.west);
\draw[dotted, line width=0.4mm] (pn2.east)--(vn02.west);
\draw[dotted, line width=0.4mm] (pn3.east)--(vn03.west);
\draw[dotted, line width=0.4mm] (pn4.east)--(vn04.west);
\draw[dotted, line width=0.4mm] (pn5.east)--(vn05.west);

\def\moveX {3.2*\nodewidth};
\def\moveXA {2*\nodewidth};

\path (cn1)++(0,0) node ()[scale=0.25] {\Huge{$I_1$}};
    \path (cn2)++(0,0) node ()[scale=0.25] {\Huge{$I_2$}};
    \path (cn3)++(0,0) node ()[scale=0.25] {\Huge{$I_3$}};
    \path (cn4)++(0,0) node ()[scale=0.25] {\Huge{$I_4$}};
     \path (cn5)++(0,0) node ()[scale=0.25] {\Huge{$I_5$}};

   \path (cn11)++(0,0) node ()[scale=0.25] {\Huge{$y_1$}};
    \path (cn12)++(0,0) node ()[scale=0.25] {\Huge{$y_2$}};
    \path (cn13)++(0,0) node ()[scale=0.25] {\Huge{$y_3$}}; 
    

\end{tikzpicture}

\caption{An example of a combined graph. Nodes $x_i^{(0)}$ and  $x_i^{(1)}$ represent individual $i$ at times 0 and 1, respectively. An interaction node $I_i$ captures interactions between individual $i$ and other individuals from time $0$ to time $1$. A factor node $y_t$ represents test $t$. Factor nodes represented by dotted squares correspond to the \textit{a priori} probability distribution of the status of individuals at time $0$.} \label{fig:Comb}
\end{figure}

 \subsection{Belief Propagation on Combined Graphs}
In this scheme, assuming that the contact tracing information is available, we form the interaction graph and combine it with the Tanner graph corresponding to the testing matrix. An example of a combined graph is presented in Fig.~\ref{fig:Comb}.
We then perform a BP algorithm over the combined graph. Note that there are two sets of variable nodes in the combined graph, $\Big\{x_i^{(0)}\Big\}_{i\in[N]}$ and $\Big\{x_i^{(1)}\Big\}_{i\in[N]}$. We are interested in computing the marginals of the posterior distribution for $\Big\{x_i^{(1)}\Big\}_{i\in[N]}$. There are also three different types of factor nodes. The interaction node $I_i$ corresponds to the conditional probability that individual $i$ at time $1$ is infected or not, given the status of individuals at time $0$ who have been in contact with individual $i$. The test factor node $y_t$ corresponds to the conditional probability that the result of test $t$ is equal to a one or zero, given the status 
of neighboring individuals at time $1$. Furthermore, there are $N$ factor nodes, represented by dashed squares, that correspond to the \textit{a priori} probability that each of the individuals at time $0$ is infected or not. The combined graph in fact represents the factorization in the following joint probability mass function.

\begin{multline*}
P\Big(\mbf{x}^{(0)}=\hat{\mbf{x}}^{(0)},\mbf{x}^{(1)}=\hat{\mbf{x}}^{(1)},\mbf{y}=\mbf{u}\Big)\\
=\Bigg[\prod_{i=1}^N P\Big({x}_i^{(0)}=\hat{{x}}_i^{(0)}\Big)\Bigg]\times \Bigg[\prod_{i=1}^N P\Big({x}_i^{(1)}=\hat{{x}}_i^{(1)}\Big \vert \Big\{{x}_{i^\prime}^{(0)}=\hat{{x}}_{i^\prime}^{(0)}\Big\}_{i^{\prime}\in \mathcal{N}(I_i)}\Big)\Bigg]\times \Bigg[\prod_{t=1}^M P\Big(y_t=u_t\Big \vert \Big \{{x}_{i}^{(1)}=\hat{{x}}_{i}^{(1)}\Big\}_{i\in \mathcal{N}(t)}\Big)\Bigg],
\end{multline*}
where $\mathcal{N}(I_i)$ denotes the indices of individuals at time $0$ who are connected to interaction node $I_i$. 

As it has been mentioned before, in a loopy BP algorithm, the messages are passed iteratively from variable nodes to factor nodes and vice versa. In what follows, we show the flow of messages in one iteration. First, individuals at time $0$ send their messages to interaction nodes. We let ${\gamma}_{i \rightarrow I_j} = [\gamma_{i\rightarrow I_j}(0) \ \gamma_{i\rightarrow I_j}(1)]$ and ${\gamma}_{I_j \rightarrow i} = [\gamma_{I_j\rightarrow i}(0) \ \gamma_{I_j\rightarrow i}(1)]$ denote the message from individual $i$ at time $0$ to interaction node $I_j$ and the message from interaction node $I_j$ to individual $i$ at time $0$, respectively. It is easy to show that the messages ${\gamma}_{i \rightarrow I_j}$ can be computed as follows: 
\begin{equation}\label{eq:itm0-to-fctr}
\begin{cases}
 \gamma_{i\rightarrow I_j}(0) \propto (1-p) \displaystyle \prod_{j^{\prime}\in \mathcal{N}(i)\setminus \{j\}}\gamma_{I_{j^{\prime}}\rightarrow i}(0), \\
 \gamma_{i\rightarrow I_j}(1) \propto p \displaystyle \prod_{j^{\prime}\in \mathcal{N}(i)\setminus \{j\}}\gamma_{I_{j^\prime}\rightarrow i}(1),\\
  \end{cases}
\end{equation}
where these messages are initialized by $\gamma_{i\rightarrow I_j}(1)=1-\gamma_{i\rightarrow I_j}(0)=p$. Then, the interaction nodes send their messages to individuals at time $1$.
We denote the message from interaction node $I_j$ to individual $j$ at time $1$ and the message from individual $j$ at time $1$ to interaction node $I_j$ by ${\delta}_{I_j \rightarrow j} = [\delta_{I_j\rightarrow j}(0) \ \delta_{I_j\rightarrow j}(1)]$ and ${\delta}_{j \rightarrow I_j} = [\delta_{j\rightarrow I_j}(0) \ \delta_{j\rightarrow I_j}(1)]$, respectively. It can be shown that the message ${\delta}_{I_j \rightarrow j}$ is given by

\begin{equation}\label{eq:Intr-to-itm1}
\begin{cases}

 \delta_{I_j\rightarrow j}(0)  \propto  \gamma_{j\rightarrow I_j}(0)  \displaystyle \prod_{i\in \mathcal{N}(I_j)\setminus \{j\}}\Big(1-q\gamma_{i\rightarrow I_j}(1)\Big ),\\
     
 \delta_{I_j\rightarrow j}(1)  \propto \gamma_{j\rightarrow I_j}(1) 
  - \gamma_{j\rightarrow I_j}(0) \displaystyle \sum_{\ell=1}^{\vert \mathcal{N}(I_j) \vert -1}\sum_{S\in \Big\{\mathcal{N}(I_j)\setminus \{j\}\Big\}_\ell}\Big(-q\Big)^\ell\prod_{i\in S}\gamma_{i\rightarrow I_j}(1).   

  \end{cases}
\end{equation}

In the next step, individuals at time $1$ send their messages to test factor nodes. We let ${\mu}_{i \rightarrow t} = [\mu_{i\rightarrow t}(0) \ \mu_{i\rightarrow t}(1)]$ and ${\mu}_{t\rightarrow i} = [\mu_{t \rightarrow i}(0) \ \mu_{t\rightarrow i}(1)]$ denote the message from individual $i$ at time $1$ to test $t$ and the message from test $t$ to individual $i$ at time $1$, respectively. 
\begin{equation}\label{eq:itm1-to-tst}
\begin{cases}
 \mu_{i\rightarrow t}(0) \propto \delta_{I_i\rightarrow i}(0) \displaystyle \prod_{t^{\prime}\in \mathcal{N}(i)\setminus \{t\}}\mu_{t^{\prime}\rightarrow i}(0), \\
 \mu_{i\rightarrow t}(1) \propto \delta_{I_i\rightarrow i}(1) \displaystyle \prod_{t^{\prime}\in \mathcal{N}(i)\setminus \{t\}}\mu_{t^{\prime}\rightarrow i}(1).\\
  \end{cases}
\end{equation}
Now, test nodes send their messages to individuals at time $1$. The message ${\mu}_{t \rightarrow i}$ is calculated in a similar way as in~\eqref{eq:tst-to-itm0} and~\eqref{eq:tst-to-itm1}. Next, individuals at time $1$ send their messages to interaction nodes. It can be shown that the message from individual $i$ at time $1$ to interaction node $I_i$ is given by
\begin{equation}\label{eq:itm1-to-fctr}
\begin{cases}
 \delta_{i\rightarrow I_i}(0) \propto \displaystyle \prod_{t\in \mathcal{N}(i)}\mu_{t\rightarrow i}(0), \\
 \delta_{i\rightarrow I_i}(1) \propto \displaystyle \prod_{t\in \mathcal{N}(i)}\mu_{t\rightarrow i}(1).\\
  \end{cases}
\end{equation}
 Finally, interaction nodes send their messages to individuals at time $0$. The message from interaction node $I_j$ to individual $j$ at time $0$ is given as follows.
\begin{equation}\label{eq:fctr-to-itm0j}
\begin{cases}
 \gamma_{I_j\rightarrow j}(0) \propto \delta_{j\rightarrow I_j}(0) \displaystyle \prod_{i\in \mathcal{N}(I_j)\setminus \{j\}}\Big(1-q\gamma_{i\rightarrow I_j}(1)\Big )
 - \delta_{j\rightarrow I_j}(1) \displaystyle \sum_{\ell=1}^{\vert \mathcal{N}(I_j) \vert -1}\sum_{S\in \Big\{\mathcal{N}(I_j)\setminus \{j\}\Big\}_\ell}\Big(-q\Big)^\ell\prod_{i\in S}\gamma_{i\rightarrow I_j}(1), \\
 \gamma_{I_j\rightarrow j}(1) \propto \delta_{j\rightarrow I_j}(1).\\
  \end{cases}
\end{equation}
The message from interaction node $I_j$ to individual $i$ at time $0$, where $i\in \mathcal{N}(I_j)\setminus \{j\}$, is given by
\begin{equation}\label{eq:fctr-to-itm0i}
\begin{cases}
\begin{split}
 \gamma_{I_j\rightarrow i}(0) & \propto \gamma_{j\rightarrow I_j}(1) \delta_{j\rightarrow I_j}(1) + \gamma_{j\rightarrow I_j}(0) \delta_{j\rightarrow I_j}(0) \displaystyle \prod_{i^{\prime}\in \mathcal{N}(I_j)\setminus \{i,j\}}\Big(1-q\gamma_{i^{\prime}\rightarrow I_j}(1)\Big )\\
 & - \gamma_{j\rightarrow I_j}(0) \delta_{j\rightarrow I_j}(1)\sum_{\ell=1}^{\vert \mathcal{N}(I_j) \vert -2}\sum_{S\in \Big\{\mathcal{N}(I_j)\setminus \{i,j\}\Big\}_\ell}\Big(-q\Big)^\ell\prod_{i^{\prime}\in S}\gamma_{i^{\prime}\rightarrow I_j}(1),    
\end{split}
 \\
\begin{split}
 \gamma_{I_j\rightarrow i}(1) & \propto \gamma_{j\rightarrow I_j}(1) \delta_{j\rightarrow I_j}(1) + (1-q)\gamma_{j\rightarrow I_j}(0) \delta_{j\rightarrow I_j}(0) \displaystyle \prod_{i^{\prime}\in \mathcal{N}(I_j)\setminus \{i,j\}}\Big(1-q\gamma_{i^{\prime}\rightarrow I_j}(1)\Big )\\
 & + \gamma_{j\rightarrow I_j}(0) \delta_{j\rightarrow I_j}(1)\Bigg[q-(1-q)\sum_{\ell=1}^{\vert \mathcal{N}(I_j) \vert -2}\sum_{S\in \Big\{\mathcal{N}(I_j)\setminus \{i,j\}\Big\}_\ell}\Big(-q\Big)^\ell\prod_{i^{\prime}\in S}\gamma_{i^{\prime}\rightarrow I_j}(1)\Bigg].   
\end{split}\\
  \end{cases}
\end{equation}

In the end, when the algorithm is run for a fixed number $T$ of iterations, we compute the marginals of the posterior distribution for individuals at time $1$ as follows.
\begin{equation*}
\begin{cases}
{{\phi}}\Big(x_{i}^{(1)}=0\Big) \propto  \delta_{I_i\rightarrow i}(0) \displaystyle \prod_{t\in \mathcal{N}(i)}\mu_{t\rightarrow i}(0), \\
  {{\phi}}\Big(x_{i}^{(1)}=1\Big) \propto  \delta_{I_i\rightarrow i}(1) \displaystyle \prod_{t\in \mathcal{N}(i)}\mu_{t\rightarrow i}(1). \\
  \end{cases}
\end{equation*}
The LLRs of the marginals are given by
\begin{equation}\label{eq:llr1}
\lambda_i=\ln \frac{ {{\phi}}\Big(x_{i}^{(1)}=1\Big)}{{{\phi}}\Big(x_{i}^{(1)}=0\Big)}=\ln \frac{ \delta_{I_i\rightarrow i}(1)}{ \delta_{I_i\rightarrow i}(0)} + \sum_{t\in \mathcal{N}(i)}\ln \frac{\mu_{t\rightarrow i}(1)}{\mu_{t\rightarrow i}(0)}.
\end{equation}
The interpretation of the LLRs is done in the same way that has been explained in Section~\ref{subsec:bp}. For a given threshold $\tau$, individual $i$ at time $1$ is announced infected if $\lambda_i \geq \tau$. 
Algorithm~\ref{alg:BP1} defines the belief propagation on combined graphs algorithm.
\begin{algorithm}
\caption{Belief Propagation on Combined Graphs}
\label{alg:BP1}
\label{CHalgorithm}
\begin{algorithmic}[1]
\State Initialize $\gamma_{i\rightarrow I_j}(1)=1-\gamma_{i\rightarrow I_j}(0)=p$ $\forall i\in [N], \forall j\in \mathcal{N}(i)$
\State Initialize $\mu_{t\rightarrow i}(0)=\mu_{t\rightarrow i}(1)=\frac{1}{2}$ $\forall t\in [M], \forall i\in \mathcal{N}(t)$ 
\For{$\ell=1,2,\cdots,T$ }
\State Compute $\delta_{I_i\rightarrow i}(0)$ and $\delta_{I_i\rightarrow i}(1)$ $\forall i\in [N]$ using~\eqref{eq:Intr-to-itm1}
\State Compute $\mu_{i\rightarrow t}(0)$ and $\mu_{i\rightarrow t}(1)$ $\forall i\in [N], \forall t\in \mathcal{N}(i)$ using~\eqref{eq:itm1-to-tst}
\State Compute $\mu_{t\rightarrow i}(0)$ and $\mu_{t\rightarrow i}(1)$ $\forall t\in [M], \forall i\in \mathcal{N}(t)$ using~\eqref{eq:tst-to-itm0} and~\eqref{eq:tst-to-itm1}
\State Compute $\delta_{i\rightarrow I_i}(0)$ and $\delta_{i\rightarrow I_i}(1)$ $\forall i\in [N]$ using~\eqref{eq:itm1-to-fctr}
\State Compute $\gamma_{I_j\rightarrow i}(0)$ and $\gamma_{I_j\rightarrow i}(1)$ $\forall j\in [N], \forall i\in \mathcal{N}(I_j)$ using~\eqref{eq:fctr-to-itm0j} and~\eqref{eq:fctr-to-itm0i}
\State Compute $\gamma_{i\rightarrow I_j}(0)$ and $\gamma_{i\rightarrow I_j}(1)$ $\forall i\in [N], \forall j\in \mathcal{N}(i)$ using~\eqref{eq:itm0-to-fctr}
\EndFor
\State Compute $\lambda_i$ $\forall i\in [N]$ using~\eqref{eq:llr1}
\end{algorithmic}
\end{algorithm}

\begin{example}\label{example}
\normalfont
Consider the combined graph shown in Fig.~\ref{fig:Comb}. We want to compute the BP messages exchanged over the edges of the combined graph. Calculating the messages from variable nodes to factor nodes is straightforward. Thus, we intend to compute messages from factor nodes to variable nodes. Since in~\cite{barbier2020strong,2018,coja2018belief,doi:10.1080/00018732.2016.1211393} it was shown that the messages from tests to individuals at time $1$ are computed using~\eqref{eq:tst-to-itm0} and~\eqref{eq:tst-to-itm1}, we only show how to compute messages from interaction nodes to individuals at time $0$ and time $1$.
The message from interaction node $I_j$ to individual $j$ at time $1$ is computed using
\[  \delta_{I_j\rightarrow j}\Big(\hat{{x}}_j^{(1)}\Big)   =\sum_{\hat{{x}}_i^{(0)}\in\{0,1\},~i\in \mathcal{N}(I_j)} \Bigg[P\Big({x}_j^{(1)}=\hat{{x}}_j^{(1)}\Big\vert \Big\{{x}_{i}^{(0)}=\hat{{x}}_{i}^{(0)}\Big\}_{i\in \mathcal{N}(I_j)}\Big)\prod_{i\in \mathcal{N}(I_j)}\gamma_{i\rightarrow I_j}\Big(\hat{{x}}_i^{(0)}\Big)\Bigg].
 \]
For instance, the message from interaction node $I_1$ to individual $1$ at time $1$ is given by 
\begin{equation}\label{eq:intr1-to-itm1}
  \delta_{I_1\rightarrow 1}\Big(\hat{{x}}_1^{(1)}\Big)   =\sum_{\hat{{x}}_1^{(0)},\hat{{x}}_3^{(0)},\hat{{x}}_5^{(0)}\in\{0,1\}} P\Big({x}_1^{(1)}=\hat{{x}}_1^{(1)}\Big\vert {x}_{1}^{(0)}=\hat{{x}}_1^{(0)},{x}_{3}^{(0)}=\hat{{x}}_3^{(0)},{x}_{5}^{(0)}=\hat{{x}}_5^{(0)}\Big)\gamma_{1\rightarrow I_1}\Big(\hat{{x}}_1^{(0)}\Big)\gamma_{3\rightarrow I_1}\Big(\hat{{x}}_3^{(0)}\Big)\gamma_{5\rightarrow I_1}\Big(\hat{{x}}_5^{(0)}\Big).
\end{equation}
 We first consider the case that $\hat{{x}}_1^{(1)}=0$, and form Table~\ref{table:1}. It is easy to see that~\eqref{eq:intr1-to-itm1} can be expanded into the following
 \begin{table*}[h]
\centering
\begin{tabular}{|c|c|c|c|}
\hline
$\hat{{x}}_1^{(0)}$ & $\hat{{x}}_3^{(0)}$ & $\hat{{x}}_5^{(0)}$ &  $P\Big({x}_1^{(1)}=0\Big\vert {x}_{1}^{(0)}=\hat{{x}}_1^{(0)},{x}_{3}^{(0)}=\hat{{x}}_3^{(0)},{x}_{5}^{(0)}=\hat{{x}}_5^{(0)}\Big)$\\ \hline
0 & 0 & 0 & 1 \\ \hline
0 & 0 & 1 & $1-q$ \\ \hline
0 & 1 & 0 & $1-q$ \\ \hline
0 & 1 & 1 & $(1-q)^2$ \\ \hline
1 & 0 & 0 & 0 \\ \hline
1 & 0 & 1 & 0 \\ \hline
1 & 1 & 0 & 0 \\ \hline
1 & 1 & 1 & 0 \\ \hline
\end{tabular}
\caption{}\label{table:1} 
\end{table*}

\[
\begin{split}
\delta_{I_1\rightarrow 1}(0) & =\gamma_{1\rightarrow I_1}(0)\gamma_{3\rightarrow I_1}(0)\gamma_{5\rightarrow I_1}(0) +
(1-q)\gamma_{1\rightarrow I_1}(0)\gamma_{3\rightarrow I_1}(0)\gamma_{5\rightarrow I_1}(1)\\
& + (1-q)\gamma_{1\rightarrow I_1}(0)\gamma_{3\rightarrow I_1}(1)\gamma_{5\rightarrow I_1}(0) + (1-q)^2\gamma_{1\rightarrow I_1}(0)\gamma_{3\rightarrow I_1}(1)\gamma_{5\rightarrow I_1}(1),
\end{split} 
\]
where it can be simplified using the fact that messages $\gamma_{i\rightarrow I_j}$ are probability distributions on $\{0,1\}$, i.e., $\mu_{i\rightarrow I_j}(0) + \mu_{i\rightarrow I_j}(1)=1$.
\[ 
\delta_{I_1\rightarrow 1}(0)= \gamma_{1\rightarrow I_1}(0) \Big(1-q\gamma_{3\rightarrow I_1}(1)\Big)\Big(1-q\gamma_{5\rightarrow I_1}(1)\Big)
\]
We now consider the case that $\hat{{x}}_1^{(1)}=1$. Expansion of~\eqref{eq:intr1-to-itm1} results in
\begin{multline*}
\delta_{I_1\rightarrow 1}(1)  =
q\gamma_{1\rightarrow I_1}(0)\gamma_{3\rightarrow I_1}(0)\gamma_{5\rightarrow I_1}(1) + q\gamma_{1\rightarrow I_1}(0)\gamma_{3\rightarrow I_1}(1)\gamma_{5\rightarrow I_1}(0) + \Big(1-(1-q)^2\Big)\gamma_{1\rightarrow I_1}(0)\gamma_{3\rightarrow I_1}(1)\gamma_{5\rightarrow I_1}(1)+\\
 \gamma_{1\rightarrow I_1}(1)\gamma_{3\rightarrow I_1}(0)\gamma_{5\rightarrow I_1}(0) + \gamma_{1\rightarrow I_1}(1)\gamma_{3\rightarrow I_1}(0)\gamma_{5\rightarrow I_1}(1) + q\gamma_{1\rightarrow I_1}(1)\gamma_{3\rightarrow I_1}(1)\gamma_{5\rightarrow I_1}(0) + q\gamma_{1\rightarrow I_1}(1)\gamma_{3\rightarrow I_1}(1)\gamma_{5\rightarrow I_1}(1),
\end{multline*}
where we can simplify it to
\[
\delta_{I_1\rightarrow 1}(1)  = \gamma_{1\rightarrow I_1}(1) - \gamma_{1\rightarrow I_1}(0)\Big(-q\gamma_{3\rightarrow I_1}(1)-q\gamma_{5\rightarrow I_1}(1)+q^2\gamma_{3\rightarrow I_1}(1)\gamma_{5\rightarrow I_1}(1) \Big).
\]

The messages from interaction nodes to individuals at time $0$ are computed as follows. The message from interaction node $I_j$ to individual $i$ at time $0$ is computed using
\[  \gamma_{I_j\rightarrow i}\Big(\hat{{x}}_i^{(0)}\Big)   =\sum_{\hat{{x}}_j^{(1)},\hat{{x}}_{i^{\prime}}^{(0)}\in\{0,1\}} P\Big({x}_j^{(1)}=\hat{{x}}_j^{(1)}\Big\vert {x}_{i}^{(0)}=\hat{{x}}_{i}^{(0)}, \Big\{{x}_{i^{\prime}}^{(0)}=\hat{{x}}_{i^{\prime}}^{(0)}\Big\}_{i^{\prime}\in \mathcal{N}(I_j)\setminus\{i\}}\Big)\delta_{j\rightarrow I_j}\Big(\hat{{x}}_{j}^{(1)}\Big)\prod_{i^{\prime}\in \mathcal{N}(I_j)\setminus\{i\}}\gamma_{i^{\prime}\rightarrow I_j}\Big(\hat{{x}}_{i^{\prime}}^{(0)}\Big).
 \]
For instance, the message from interaction node $I_1$ to individual $3$ at time $0$ is given by 
\begin{equation}\label{eq:intr1-to-itm3}
 \gamma_{I_1\rightarrow 3}\Big(\hat{{x}}_{3}^{(0)}\Big)   =\sum_{\hat{{x}}_{1}^{(1)},\hat{{x}}_{1}^{(0)},\hat{{x}}_{5}^{(0)}\in\{0,1\}} P\Big({x}_1^{(1)}=\hat{{x}}_{1}^{(1)}\Big\vert {x}_{1}^{(0)}=\hat{{x}}_{1}^{(0)},{x}_{3}^{(0)}=\hat{{x}}_{3}^{(0)},{x}_{5}^{(0)}=\hat{{x}}_{5}^{(0)}\Big)\delta_{1\rightarrow I_1}\Big(\hat{{x}}_{1}^{(1)}\Big)\gamma_{1\rightarrow I_1}\Big(\hat{{x}}_{1}^{(0)}\Big)\gamma_{5\rightarrow I_1}\Big(\hat{{x}}_{5}^{(0)}\Big).
\end{equation} 
First, we consider the case that $\hat{{x}}_{3}^{(0)}=0$, and expand~\eqref{eq:intr1-to-itm3} as follows.
\[
\begin{split}
\gamma_{I_1\rightarrow 3}(0) & = \delta_{1\rightarrow I_1}(0)\gamma_{1\rightarrow I_1}(0)\gamma_{5\rightarrow I_1}(0) + (1-q)\delta_{1\rightarrow I_1}(0)\gamma_{1\rightarrow I_1}(0)\gamma_{5\rightarrow I_1}(1) + q\delta_{1\rightarrow I_1}(1)\gamma_{1\rightarrow I_1}(0)\gamma_{5\rightarrow I_1}(1)\\
& + \delta_{1\rightarrow I_1}(1)\gamma_{1\rightarrow I_1}(1)\gamma_{5\rightarrow I_1}(0) + \delta_{1\rightarrow I_1}(1)\gamma_{1\rightarrow I_1}(1)\gamma_{5\rightarrow I_1}(1),
\end{split}
\]
where after simplification becomes
\[
\gamma_{I_1\rightarrow 3}(0)= \gamma_{1\rightarrow I_1}(1)\delta_{1\rightarrow I_1}(1)+ \gamma_{1\rightarrow I_1}(0)\delta_{1\rightarrow I_1}(0)\Big(1- q\gamma_{5\rightarrow I_1}(0)\Big)-\gamma_{1\rightarrow I_1}(0)\delta_{1\rightarrow I_1}(1)\Big(-q \gamma_{5\rightarrow I_1}(1)\Big).
\]
Then, we consider the case that $\hat{{x}}_{3}^{(0)}=1$. 
\[
\begin{split}
\gamma_{I_1\rightarrow 3}(1) & = (1-q)\delta_{1\rightarrow I_1}(0)\gamma_{1\rightarrow I_1}(0)\gamma_{5\rightarrow I_1}(0) + (1-q)^2\delta_{1\rightarrow I_1}(0)\gamma_{1\rightarrow I_1}(0)\gamma_{5\rightarrow I_1}(1) + q\delta_{1\rightarrow I_1}(1)\gamma_{1\rightarrow I_1}(0)\gamma_{5\rightarrow I_1}(0)\\
& + \Big(1- (1-q)^2\Big)\delta_{1\rightarrow I_1}(1)\gamma_{1\rightarrow I_1}(0)\gamma_{5\rightarrow I_1}(1) + \delta_{1\rightarrow I_1}(1)\gamma_{1\rightarrow I_1}(1)\gamma_{5\rightarrow I_1}(0) + \delta_{1\rightarrow I_1}(1)\gamma_{1\rightarrow I_1}(1)\gamma_{5\rightarrow I_1}(1),
\end{split}
\]
where can be simplified to
\[
\begin{split}
\gamma_{I_1\rightarrow 3}(1) &= \gamma_{1\rightarrow I_1}(1)\delta_{1\rightarrow I_1}(1)+ (1-q)\gamma_{1\rightarrow I_1}(0)\delta_{1\rightarrow I_1}(0)\Big(1- q\gamma_{5\rightarrow I_1}(0)\Big)\\
&+\gamma_{1\rightarrow I_1}(0)\delta_{1\rightarrow I_1}(1)\Bigg[q-(1-q)\Big(-q \gamma_{5\rightarrow I_1}(1)\Big)\Bigg].
\end{split} 
\]  \hfill \qedsymbol{}
\end{example}

\begin{figure*}[tb]
\begin{subfigure}{.49\textwidth}\centering \includegraphics[width=.95\linewidth]{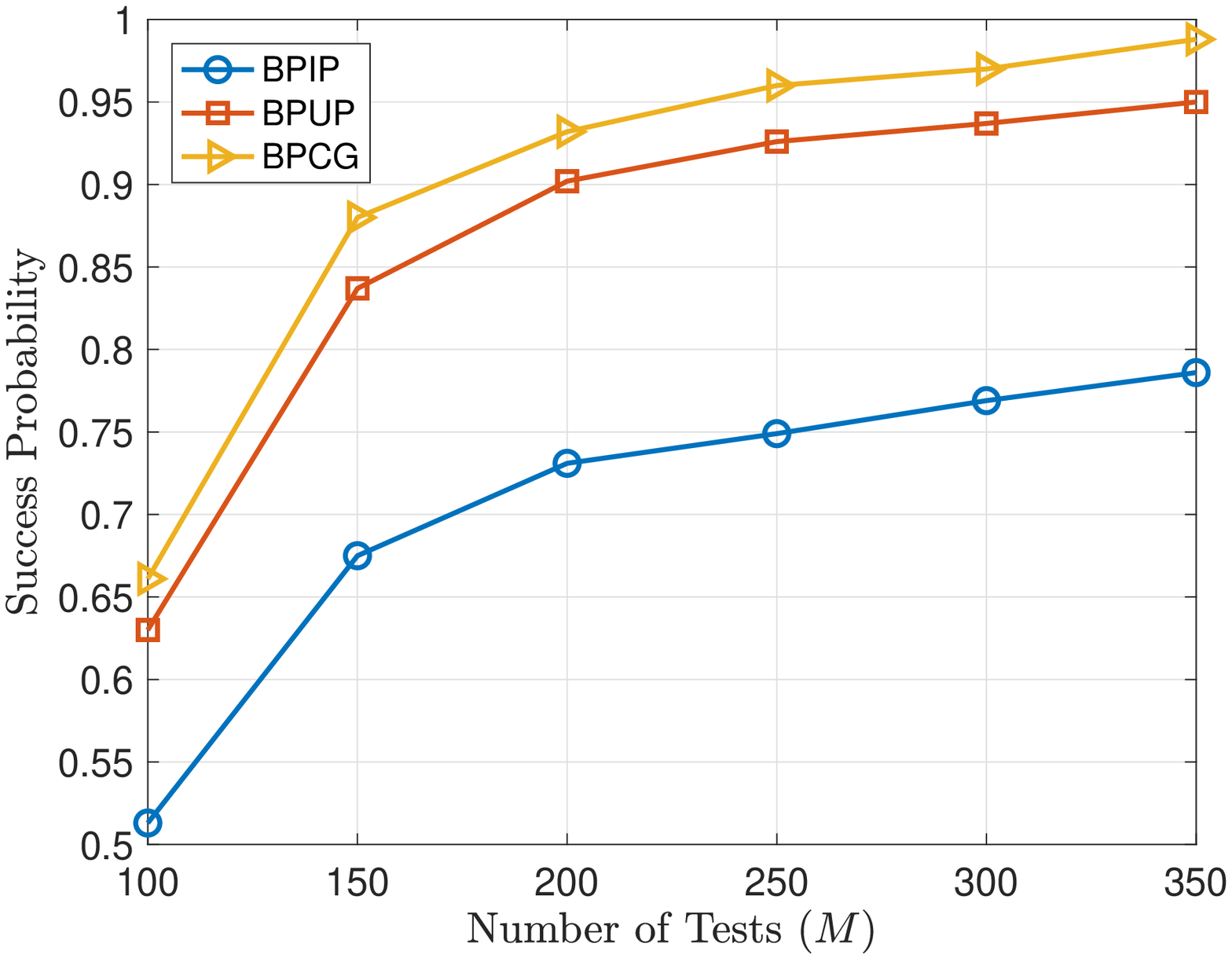}  
  \caption{$\rho=0.01$}
  \label{fig:sub-second}
\end{subfigure}
\begin{subfigure}{.49\textwidth}
  \centering
  \includegraphics[width=.95\linewidth]{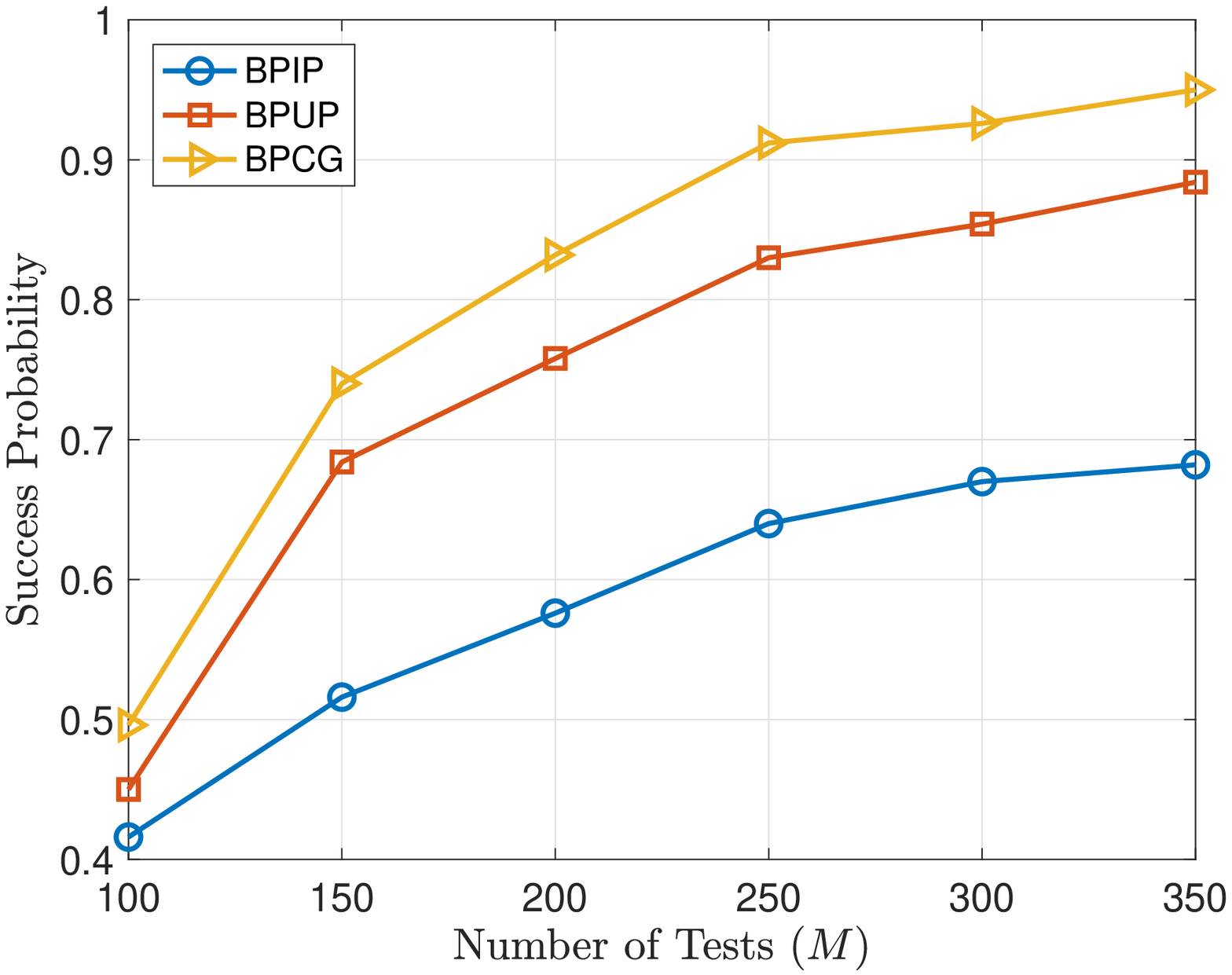}  
  \caption{$\rho=0.05$}
  \label{fig:sub-fourth}
\end{subfigure}
\caption{Success probability as a function of the number of tests $M$ based on simulation results for $N=500$ individuals, the prevalence rate $p=0.01$, the contagion probability $q=0.1$, and the interaction probability $\theta=0.008$, under the binary symmetric noise model with parameter $\rho\in\{0.01,0.05\}$
.}
\label{fig:sp}\vspace{-0.25cm}
\end{figure*}

\section{Simulation Results}\label{sec:SR}
In this section, we compare the performance of the BP using Initial Prior probabilities (BPIP) algorithm, the BP using Updated Prior probabilities (BPUP) algorithm, and the BP on Combined Graphs (BPCG) algorithm using three metrics, success probability, FNR, and FPR. 
Each result is averaged over $1000$ experiments. The testing matrix is constructed according to a Bernoulli design with parameters $\nu=\ln 2$. In the BPUP algorithm, the updated prior probabilities given by~\eqref{eq:prior} are computed using the contact tracing information. In our simulations, we assume that individual $i$ at time $0$, for each $i\in [N]$, interacts with individual $j$ at time $0$, for each $j\in [N]\setminus \{i\}$, with some fixed probability $\theta$, referred to as \emph{interaction probability}. It should be noted that $d_i$, the number of individuals which interact with individual $i$, follows a binomial distribution with parameters $N-1$ and $\theta$, i.e., $d_i\sim B(N-1,\theta)$. The expected value of $x_i^{(1)}$ is computed as follows:
\[\mathbb{E}\Big[x_i^{(1)}\Big]= \mathbb{E}\Big[ \mathbb{E}\Big[x_i^{(1)}\Big\vert d_i\Big] \Big] = \mathbb{E}\Big[1-(1-p)(1-pq)^{d_i}\Big]=1-(1-p)\mathbb{E}\Big[(1-pq)^{d_i}\Big],\]
where the term $\mathbb{E}\Big[(1-pq)^{d_i}\Big]$ is given by
\[
\mathbb{E}\Big[(1-pq)^{d_i}\Big]=\sum_{d=0}^{N-1}(1-pq)^d {N-1\choose d} \theta^{d}(1-\theta)^{N-d-1}=(1-pq\theta)^{N-1}.
\]
Thus, we have $\mathbb{E}\Big[x_i^{(1)}\Big]=1-(1-p)(1-pq\theta)^{N-1}$. Accordingly, the average number of infected individuals for the BPUP and the BPCG algorithms is given by $K=N\Big(1-(1-p)(1-pq\theta)^{N-1}\Big)$.

In Fig.~\ref{fig:sp}, we plot success probability as a function of the number of tests $M$ based on simulation results for $N=500$ individuals, the prevalence rate $p=0.01$, the contagion probability $q=0.1$, and the interaction probability $\theta=0.008$, under the binary symmetric noise model with parameter $\rho\in\{0.01,0.05\}$. The value of success probability for each number of test is optimized over the threshold in the range $\tau\in[-10,10]$. The number of iterations for the BPUP and the BPIP algorithms is $T=15$. We consider $T=30$ iterations for the BPCG algorithm.
It can be observed that the BPCG algorithm outperforms the other algorithms for all values of $M$. For instance, when $\rho=0.01$ and the number of test is $M=350$, the BPCG algorithm provides a success probability $4\%$ and $24\%$ greater than that of the BPUP and the BPIP algorithms, respectively.
Also, it can be seen that for the high noise regime, i.e., $\rho=0.05$, the advantage of BPCG algorithm over the other algorithms in terms of success probability becomes more evident. For example, for $\rho=0.05$ and $M=350$ tests, the success probability of the BPCG algorithm is $7.4\%$ and $43.5\%$ greater than that of the BPUP and the BPIP algorithms, respectively. 

\begin{figure*}[tb] \vspace{-0.1cm}
\begin{subfigure}{.49\textwidth}\centering \includegraphics[width=.95\linewidth]{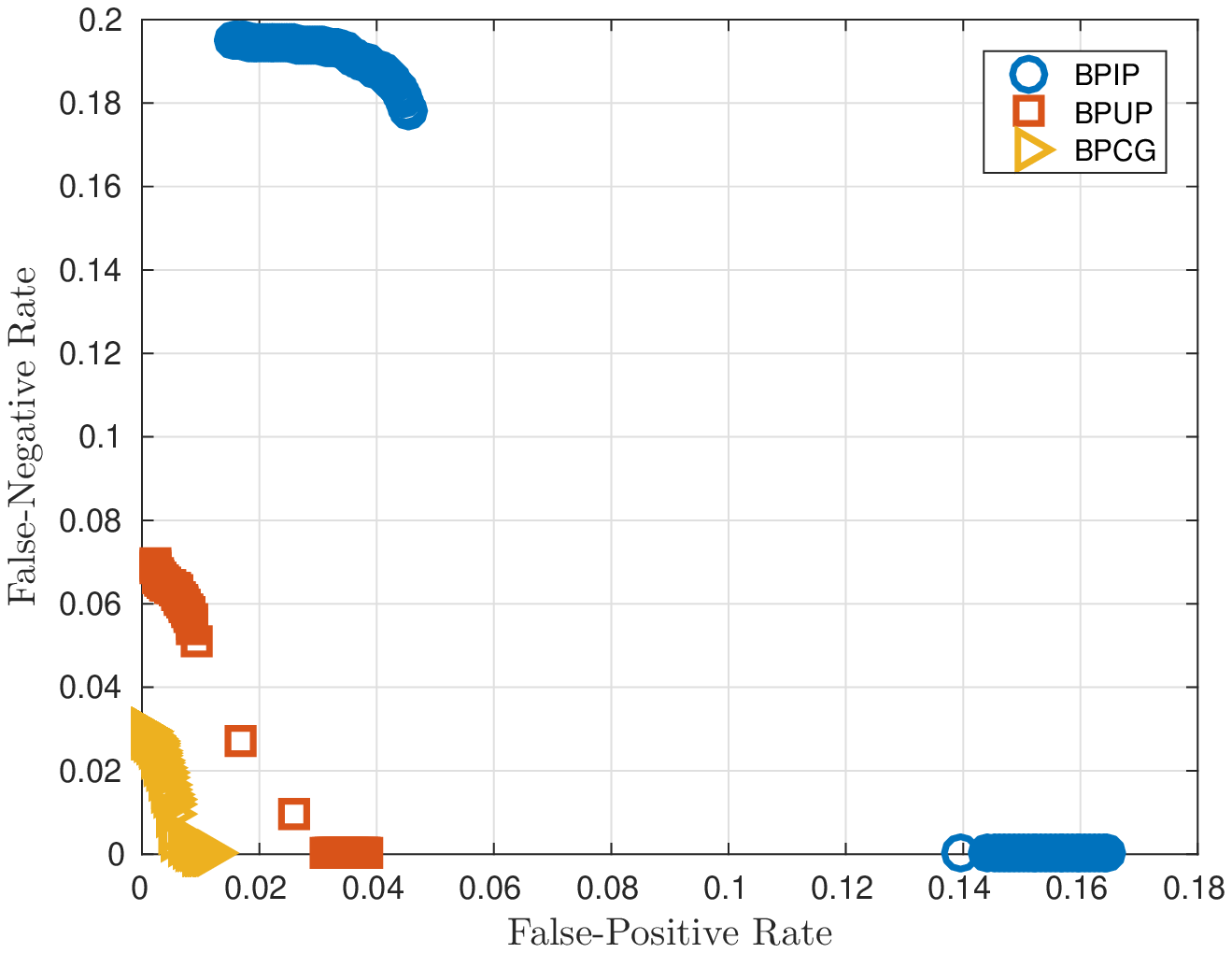}  
  \caption{$\rho=0.01$}
  \label{fig:sub-second}
\end{subfigure}
\begin{subfigure}{.49\textwidth} \vspace{-0.1cm}
  \centering
  \includegraphics[width=.95\linewidth]{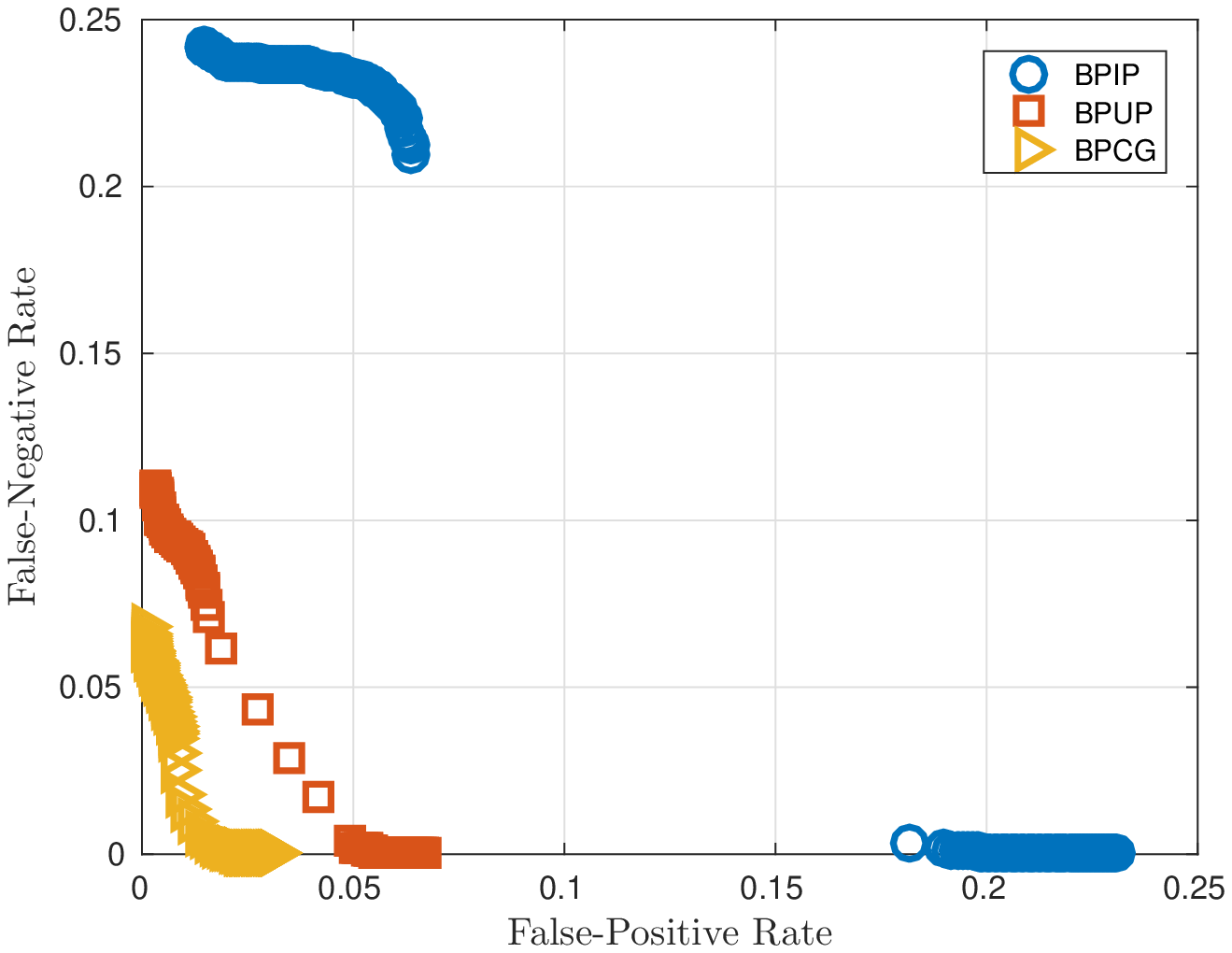}  
  \caption{$\rho=0.05$}
  \label{fig:sub-fourth}
\end{subfigure}
\caption{FNR vs. FPR based on simulation results for threshold $\tau\in[-10,10]$, $N=500$ individuals, $M=350$ tests, the prevalence rate $p=0.01$, the contagion probability $q=0.1$, and the interaction probability $\theta=0.008$, under the binary symmetric noise model with parameter $\rho\in\{0.01,0.05\}$
.} \vspace{-0.25cm}
\label{fig:fnrfpr}
\end{figure*}

 In Fig.~\ref{fig:fnrfpr}, we depict FNR vs. FPR for all three decoding algorithms for threshold $\tau\in[-10,10]$, $N=500$ individuals, $M=350$ tests, the prevalence rate $p=0.01$, the contagion probability $q=0.1$, and the interaction probability $\theta=0.008$, under the binary symmetric noise model with parameter $\rho\in\{0.01,0.05\}$. Note that unlike the success probability, FNR and FPR do not converge. Instead, after a certain number of iterations, FNR and FPR oscillate around an average value. Thus, for each value of $\tau$, we compute the average FNR and FPR over a range of iterations. The range of iterations for the BPUP and the BPIP algorithms is $T\in\{15,16,\cdots,30\}$. For the BPCG algorithm, we consider the range of iterations $T\in\{30,31,\cdots,50\}$  .
 Each point on the curve corresponding to a decoding algorithm represents the pair (FNR,FPR) which has been computed for the same value of $\tau$. The closer a curve is to the origin of the FNR–FPR plane, the better the performance of the corresponding decoding algorithm in terms of FNR and FPR. It can be observed that for the BPCG algorithm the operating point that minimizes the total error rate, i.e., the sum of FPR and FNR, is closer to the origin than that of the BPUP and the BPIP algorithms.

\vspace{-0.17cm}

\bibliographystyle{IEEEtran}
\bibliography{NGTRefs}

\end{document}